\documentclass[10pt,aps,pra,reprint,superscriptaddress]{revtex4-1}
\pdfoutput=1

\usepackage[utf8]{inputenc} 
\usepackage[T1]{fontenc}
\usepackage{amsmath}        %
\usepackage{amssymb}
\usepackage{amsfonts}
\usepackage{graphicx}       %
\usepackage{esdiff} 

\usepackage{dsfont}         
\usepackage{braket}         

\usepackage[hidelinks]{hyperref}       

\usepackage{tikz}

\newcommand{\LR}[1]{\left(#1\right)}                   
\newcommand{\Abs}[1]{\left\vert#1\right\vert}
\newcommand{\ii}{\mathrm{i}}                           

\newcommand{\Trace}[1]{\text{tr}\left\{#1\right\}}     

\begin{document}

\title{Implementation of the iFREDKIN gate in scalable superconducting architecture for the quantum simulation of Fermionic systems}
\author{Per J. Liebermann}
\affiliation{Theoretical physics, Saarland University, 66123 Saarbr\"ucken, Germany}
\author{Pierre-Luc Dallaire-Demers}
\affiliation{Department of Chemistry and Chemical Biology, Harvard University, Cambridge, MA 02138, USA}
\author{Frank K. Wilhelm}
\affiliation{Theoretical physics, Saarland University, 66123 Saarbr\"ucken, Germany}
\date{\today}

\begin{abstract}
	We present a Superconducting Planar ARchitecture for Quantum Simulations (SPARQS) intended to implement a scalable qubit layout for quantum simulators. 	To this end, we describe the  iFREDKIN gate as a controlled entangler for the simulation of Fermionic systems that is advantageous if it can be directly implemented. Using optimal control, we show that and how this gate can be efficiently implemented in the SPARQS circuit, making it a promising platform and control scheme for quantum simulations. Such a quantum simulator can be built with current quantum technologies to advance the design of molecules and quantum materials.
\end{abstract}

\maketitle

\section{\label{sec:intro}Introduction} 
Richard Feynman proposed building computers that processes information based on the rules of quantum mechanics to solve the difficult problem of simulating quantum systems \cite{Feynman1982}.
Motivated by Shor's algorithm to factor large composite numbers \cite{Shor1999} and its potential in cryptography, considerable efforts have been invested in building a universal fault-tolerant quantum computer \cite{DiVincenzo2000,Ladd2010,Fujii2015}. 
However, it was 
recently argued that such a universal machine may not be necessary for the purpose of simulating Fermionic systems beyond the reach of modern supercomputers 
\cite{Bloch2012,Blatt2012,AspuruGuzik2012,Houck2012,Georgescu2014} as long coherence times (quantum memories) ought to be sufficient.
Following advances in quantum technologies from the past decades, it can be suggested that Feynman's machine could possibly be built in a near future.
It is entirely possible that quantum simulations of 
Fermionic systems such as Fermi-Hubbard-like models and molecular models will be a main application of quantum computers in the coming years.

In this paper, we propose a superconducting circuit architecture for simulating general Fermionic systems which can be cast in a second-quantized formulation.
It is based on a superconducting circuit implementation called the RezQu architecture, as this has the tunability of its components and the simplicity of its implementation \cite{Mariantoni2011,Galiautdinov2012,Ghosh2013}.
The layout presented in sec.~\ref{sec:architecture} is extensible and planar and it can be implemented with current quantum technologies.
The properties of this Superconducting Planar ARchitecture for Quantum Simulations (SPARQS) are such that it can be used to prepare a molecular or cluster state, measure its energy and correlation functions \cite{DDW16}.

It was also previously shown that for quantum simulations, the number of gates that need to be tuned and benchmarked scales linearly with the size of the simulated system \cite{DDW16b}. For example, in the case of the Fermi-Hubbard model, the size of the system in a hybrid simulation method \cite{Potthoff2003,Senechal2008a,DDW16} corresponds to the number of spin orbitals in an exactly solved sub-lattice of the full infinite lattice.
From the RezQu literature, we assume that single-qubit and two-qubit gates can be implemented straightforwardly based on known results \cite{Mariantoni2011,Ghosh2013,Egger2014}. 
Quantum simulations also benefit from iFREDKIN gates to efficiently implement the time evolution of Fermionic Hamiltonians.
The iFREDKIN gate is a new entangling gate in the family of three-qubit gates which includes the TOFFOLI and FREDKIN gates. However, unlike the latter two, it has no classical analog. It performs an entangler, the iSWAP \cite{Calderon-Vargas2015}, on two target qubits conditioned on a control qubit (conditional iSWAP). This conditional evolution is naturally adapted to the need to interfere an entangled manybody state in the target qubit to a reference state as a key step in phase estimation.
We expect it to be the most costly gate in quantum simulations, as it is used between chains of qubits to implement hopping and interaction terms of Fermionic Hamiltonians \cite{DDW16b}.
Therefore, the remaining challenge is to show that the iFREDKIN gates can be implemented between the probe qubit P and neighboring system qubits S.

In sec.~\ref{sec:iFREDKIN}, as a proof of principle, we use GRadient Ascent Pulse Engineering (GRAPE) \cite{Khaneja2005,Glaser2015} 
to show that the iFREDKIN gate can be implemented in a time comparable to a simple iSWAP gate between neighboring system qubits, even when leakage is included in a SPARQS circuit. Thus, we conclude that SPARQS circuits with an appropriate iFREDKIN control scheme provide a natural platform for the simulation of Fermionic systems.

\section{\label{sec:architecture}Circuit architecture} 
\begin{figure}[h]
  \includegraphics[width=\columnwidth]{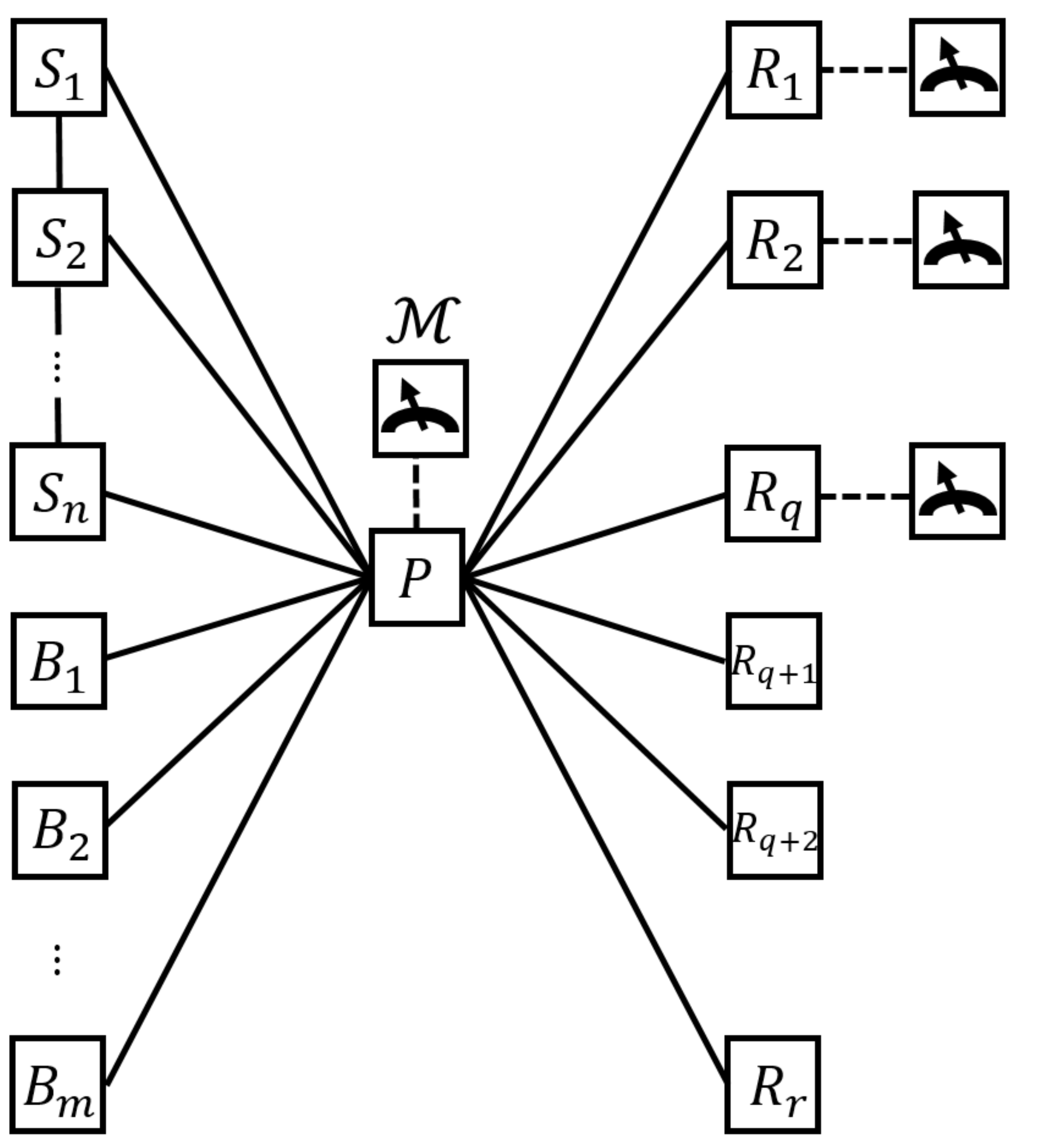}
  \caption{The $S$ register is used to encode the system dynamics, the $B$ register contains bath qubits used to prepare a simulated state in $S$.
	$P$ is used to measure the correlation functions of the simulated system and to mediate the interaction with the digital register $R$,
	which yields information on the temperature of the Gibbs state prepared in $S$. (Figure taken from \cite{DDW16b})}
	\label{fig:Layout}
\end{figure}

In Ref.~\cite{DDW16b} we highlighted that a dual-rail qubit with a highly connected central qubit is well-suited for the purpose of simulating clusters of the Fermi-Hubbard model and other Fermionic systems.
As shown in fig.~\ref{fig:Layout}, the layout consists of a register $S$ which encodes a system Hamiltonian and a bath register $B$ used in the procedure of creating a Gibbs state of the system Hamiltonian in $S$. The Gibbs state preparation \cite{Riera2012} or phase estimation \cite{Nielsen2010book} digital register $R$ also requires a line of qubits whose size depends on the desired precision of prepared or measured energies. Interactions between registers $S+B$ and $R$ and possible subsequent correlation functions measurements are mediated through a probe qubit $P$ between the digital and analog registers. Lines between registers indicates where multi-qubit interactions are used in simulation algorithms. An advantage of using a middle qubit $P$ is that all-to-all connectivity is not required for the implementation of useful algorithms. The triangles formed by interaction lines between neighboring qubits in $S$ and register $P$ are meant to indicate that iFREDKIN gates have to be used with $P$ as the control qubit. 

Here, we will describe a possible superconducting circuit architecture for this purpose. Within this, we will show in sec.~\ref{sec:iFREDKIN} how to implement a fast and direct iFREDKIN gate using optimal control methods. 

\begin{figure}[h]
  \includegraphics[width=\columnwidth]{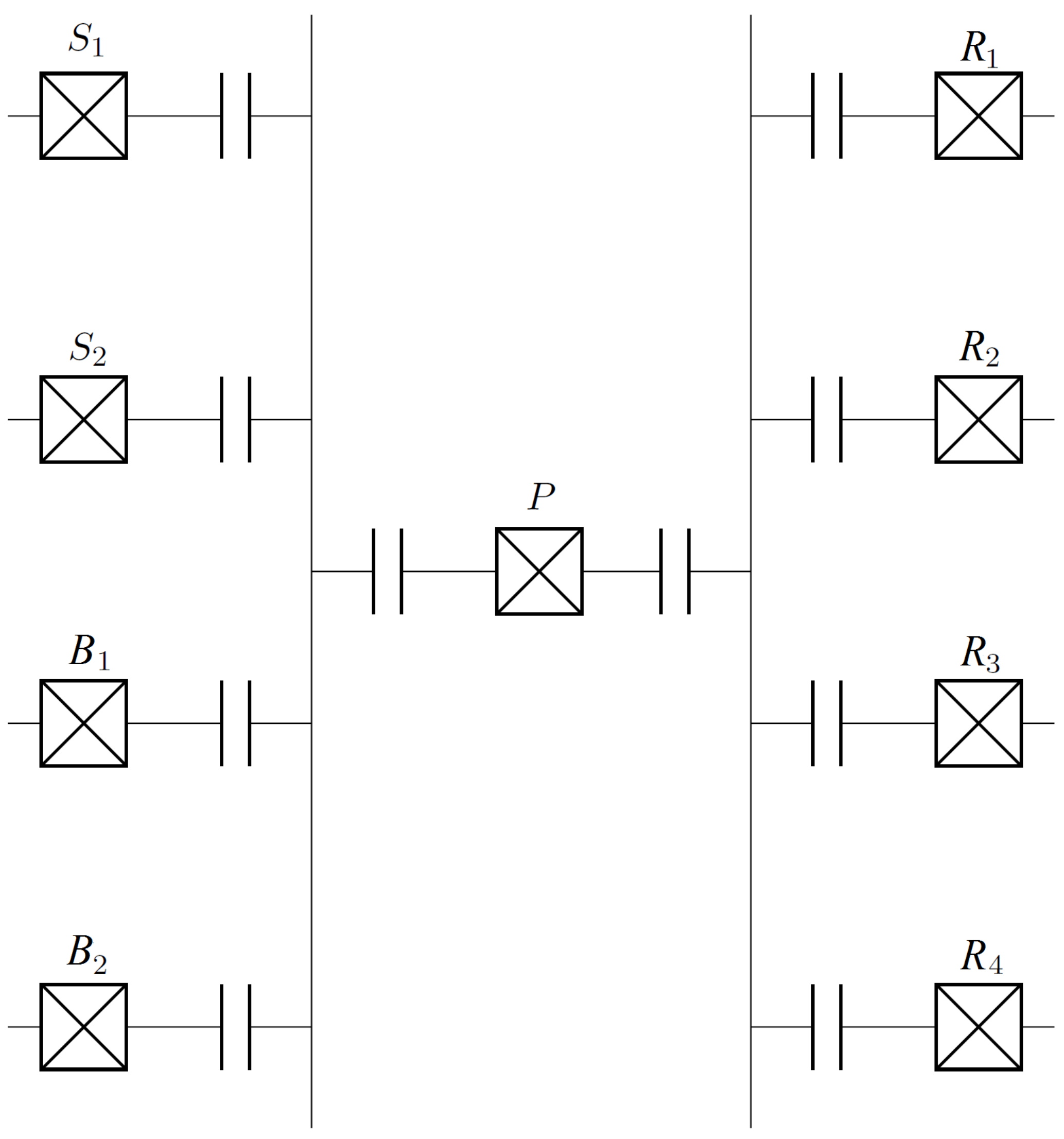}
  \caption{SPARQS: Superconducting Planar ARchitecture for Quantum Simulations. It is a modified RezQu architecture. Each frequency-tunable qubit (represented by a crossed box) is coupled to a common transmission line. Not shown are the flux control lines of the qubits to change their detuning from the bus.}
	\label{fig:SPARQS}
\end{figure}

The basic architecture for such a SPARQS circuit is shown in in fig.~\ref{fig:SPARQS}, a modified RezQu architecture \cite{Mariantoni2011,Galiautdinov2012,Ghosh2013}.
Qubits with tunable frequency are connected through a superconducting cavity which acts as a bus for quantum information \cite{Blais2004, Blais2007}.
The qubits do not interact with each other unless they are brought in resonance with the cavity. This architecture can be fabricated in a planer way and is expeced to be extensible based on current quantum technologies. The exponential increase in coherence times of superconducting qubits \cite{Geerlings13} is a good indication that this architecture could be tested with minimal quantum error correction.
It is known that single-qubit and two-qubit gate can be efficiently implemented in RezQu-like circuits \cite{Egger2014}. However, iFREDKIN gates have not been studied yet. In the next section, we show using optimal control that the iFREDKIN gate can be implemented efficiently in a SPARQS processor for realistic circuit parameters.

\section{\label{sec:iFREDKIN}iFREDKIN} 

Analogous to the FREDKIN gate (conditional swap), the iFREDKIN gate is an entangling three-qubit gate, which performs an iSWAP operation on two qubits, depending on the state of the first qubit, i.e., a conditional iSWAP:
\begin{align}
U_{\pm\mathrm{iFREDKIN}}&=\left|0\right\rangle \left\langle 0\right|\otimes\mathds{1}_4+\left|1\right\rangle \left\langle 1\right|\otimes\pm\mathrm{iSWAP}\nonumber\\
&=\begin{bmatrix}
1 & 0 & 0 & 0 & 0 & 0 & 0 & 0 \\
0 & 1 & 0 & 0 & 0 & 0 & 0 & 0 \\
0 & 0 & 1 & 0 & 0 & 0 & 0 & 0 \\
0 & 0 & 0 & 1 & 0 & 0 & 0 & 0 \\
0 & 0 & 0 & 0 & 1 & 0 & 0 & 0 \\
0 & 0 & 0 & 0 & 0 & 0 & \pm\ii & 0 \\
0 & 0 & 0 & 0 & 0 & \pm\ii & 0 & 0 \\
0 & 0 & 0 & 0 & 0 & 0 & 0 & 1 \\
\end{bmatrix}\,.
\end{align}
It is used in the context of quantum simulations to perform the time evolution of Fermionic Hamiltonians \cite{DDW16b} in a Jordan-Wigner basis \cite{Jordan1928}, which maps indistinguishable particles with antisymmetric exchange properties to a register of distinguishable qubits. The iFREDKIN gate is entangling since it will map a separable state $\left(\ket{001}+\ket{101}\right)/\sqrt{2}$ to a generalized GHZ state $\left(\ket{001}+i\ket{110}\right)/\sqrt{2}$ \cite{Dur2000}. Specifically, it executes a two-qubit iSWAP gate, which is a perfect two-qubit entangler \cite{Zhang2003}, conditional on a control qubit. When the control qubit is in a superposition, it accumulates a phase $i$ if and only if the state of the system qubits are different. Hence, the iFREDKIN gate can be used to characterize the interaction between qubits as if they where indistinguishable particles with antisymmetric exchange properties. In a SPARQS circuit, the control qubit is always $P$ and the conditional iSWAP is performed between neighboring $S$ qubits.

\begin{figure}[h]
  \includegraphics[width=2in]{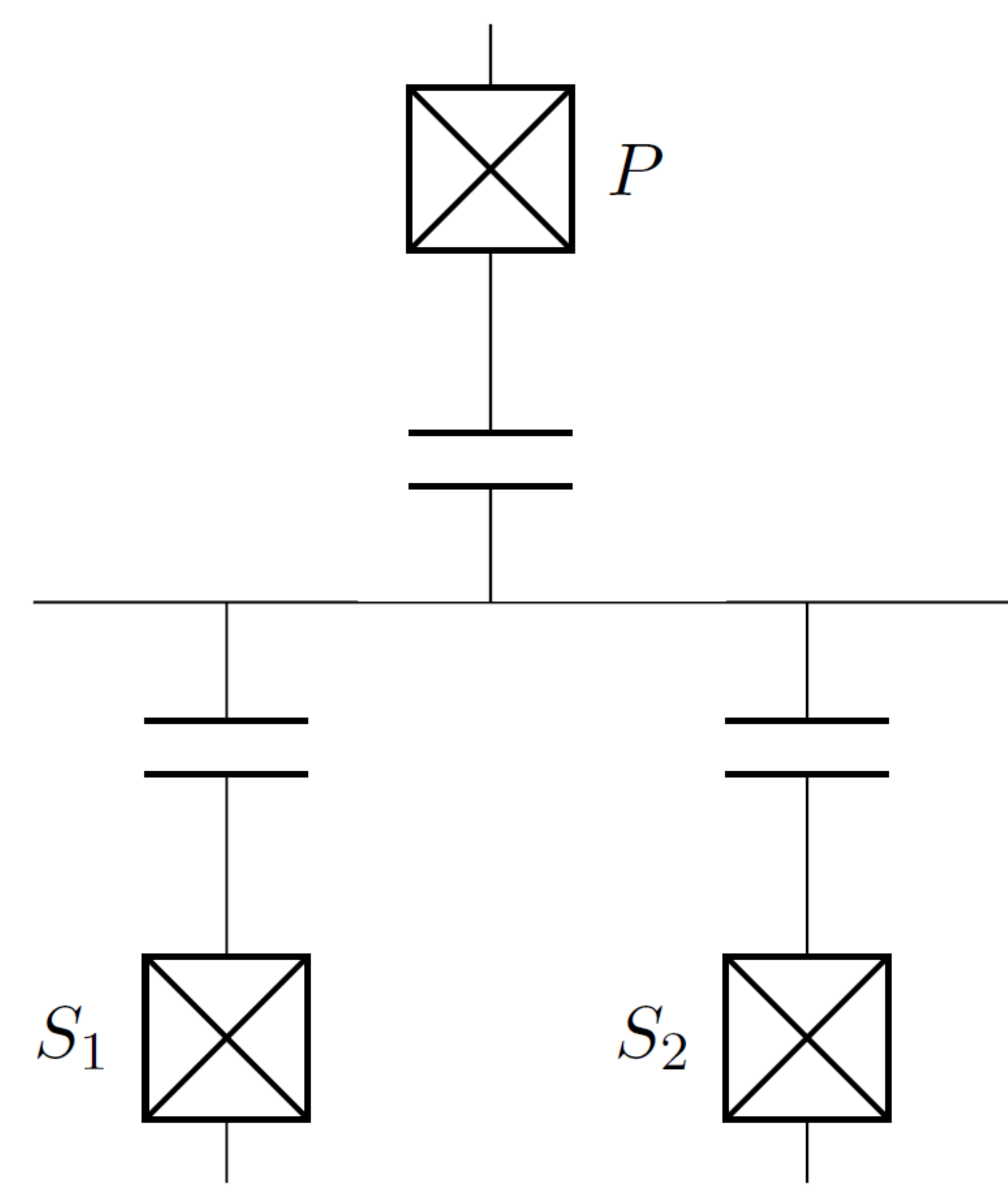}
  \caption{Unit cell for the optimal control problem. It is assumed that the other qubits are decoupled from interacting with each other during the control sequence by detuning them from the bus.}
	\label{fig:UnitCell}
\end{figure}
Here we want to implement the iFREDKIN gate on a cluster of the SPARQS processor as shown in fig.~\ref{fig:UnitCell}. Pulse shapes found by numerical methods, such as GRAPE \cite{Khaneja2005}, have proven to be faster than analytical control pulses on this architecture \cite{Egger2014}, and optimal control methods have been demonstrated successfully on three qubit gates \cite{Zahedinejad2015}. Additionally, we avoid to reach the decoherence limit compared to a gate decomposition of multi-qubit gates.

\begin{figure}[h]
  \includegraphics[width=\columnwidth]{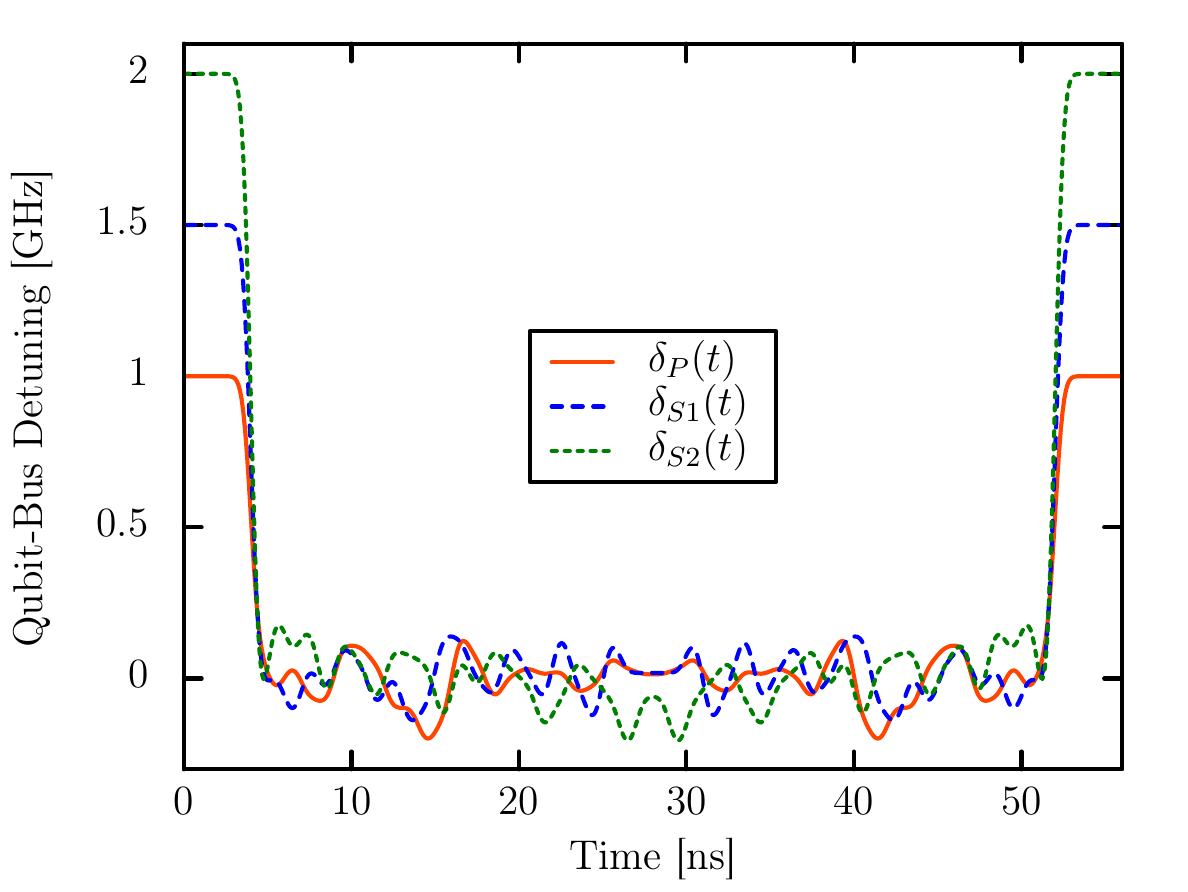}
  \caption{The 3 $Z$-controls, including a 4\,ns buffer on each side and a Gaussian filter with standard deviation $\sigma=0.4$\,ns. The coupling between the qubits is mediated by a bus.}
  \label{fig:zctrls}
\end{figure}

\begin{figure}[hb]
  \includegraphics[width=\columnwidth]{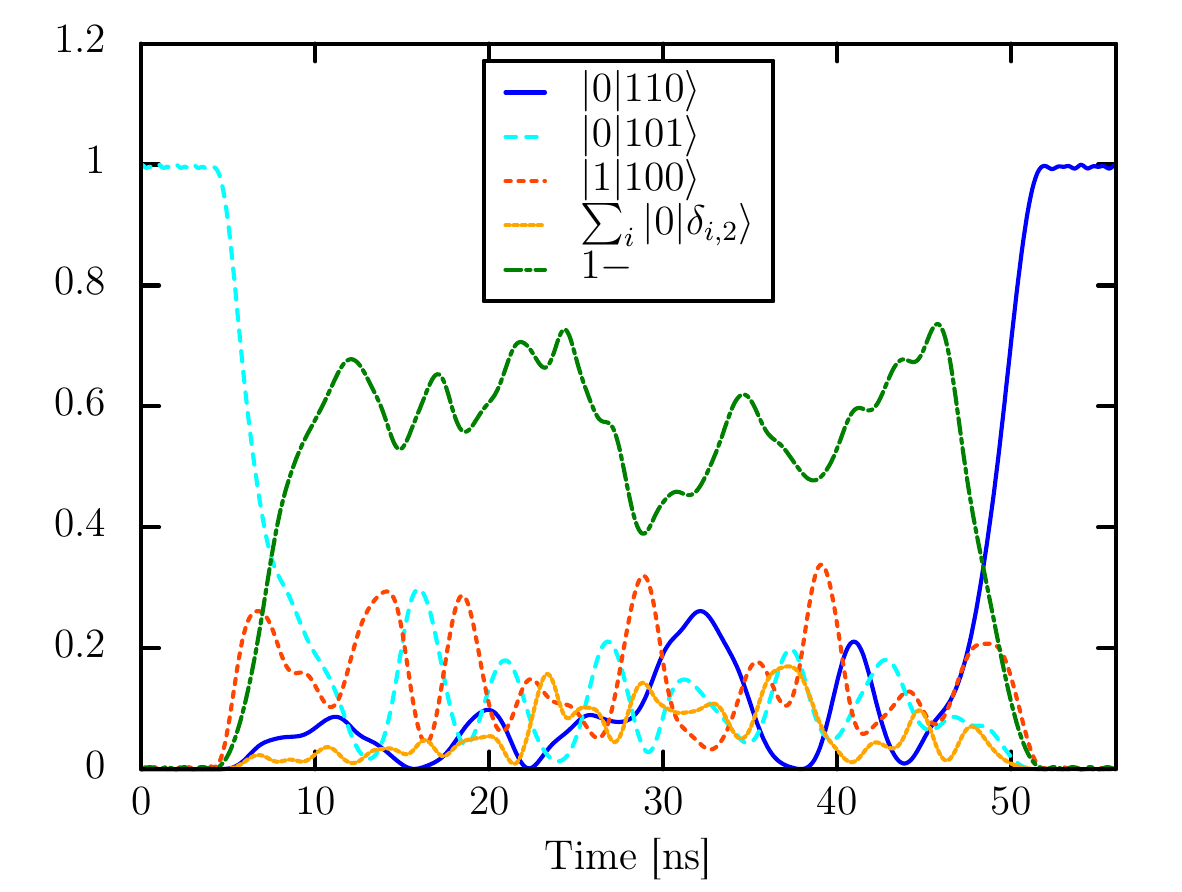}\\
  \includegraphics[width=\columnwidth]{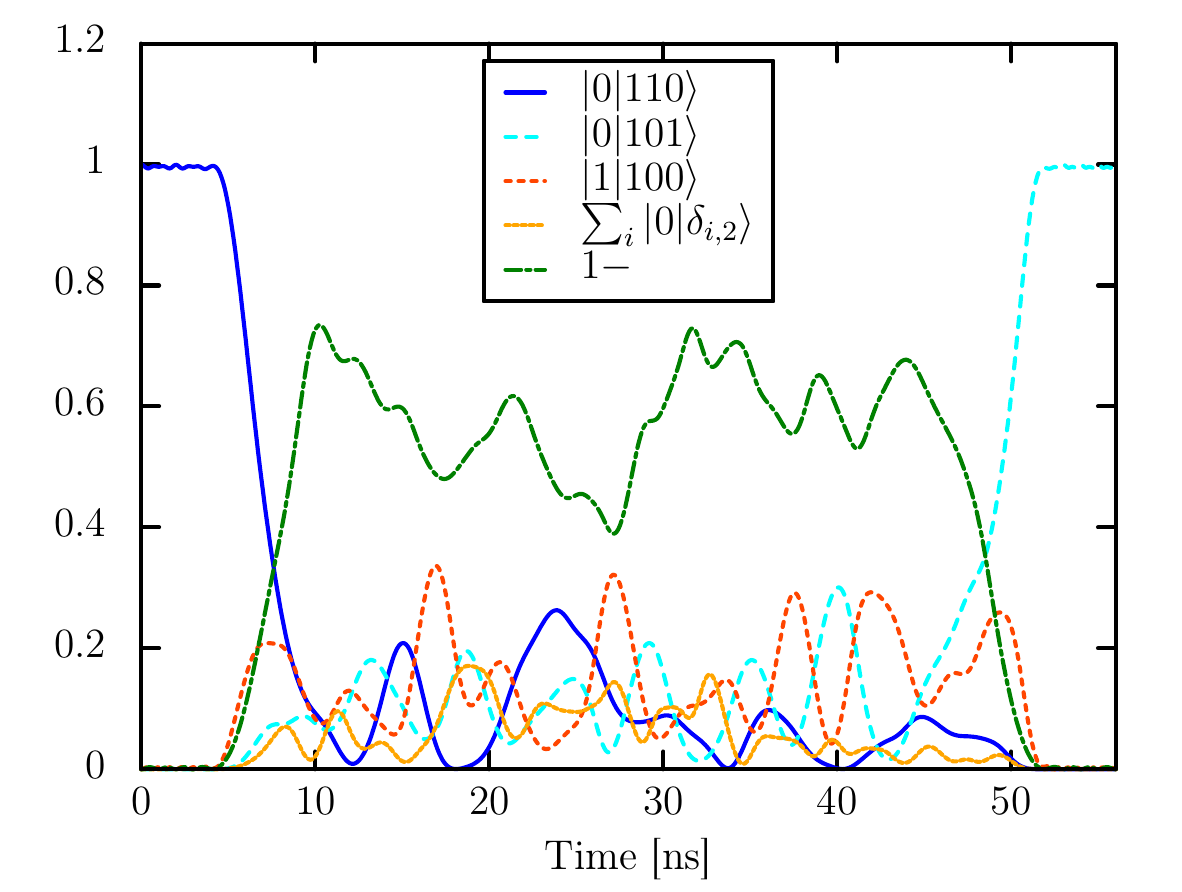}
  \caption{Time evolution in the subspace with two excitations. Shown are the population of the states $\ket{0|110}$ and $\ket{0|101}$, which are swapped, the population of the intermediate state $\ket{1|100}$, the sum of populations of the leakage levels $\sum_i\ket{0|\delta_{i,2}}=\ket{0200}+\ket{0020}+\ket{0002}$ and the sum of populations in the four other states $1- = \ket{0|011}+\ket{1|010}+\ket{1|001}+\ket{2|000}$. There is some excursion in the leakage levels which is canceled by the end of the pulse. The inclusion of this extra Hilbert space may be beneficial for the efficiency of the gate.}
  \label{fig:pop}
\end{figure}

\begin{figure}[t]
  \includegraphics[width=\columnwidth]{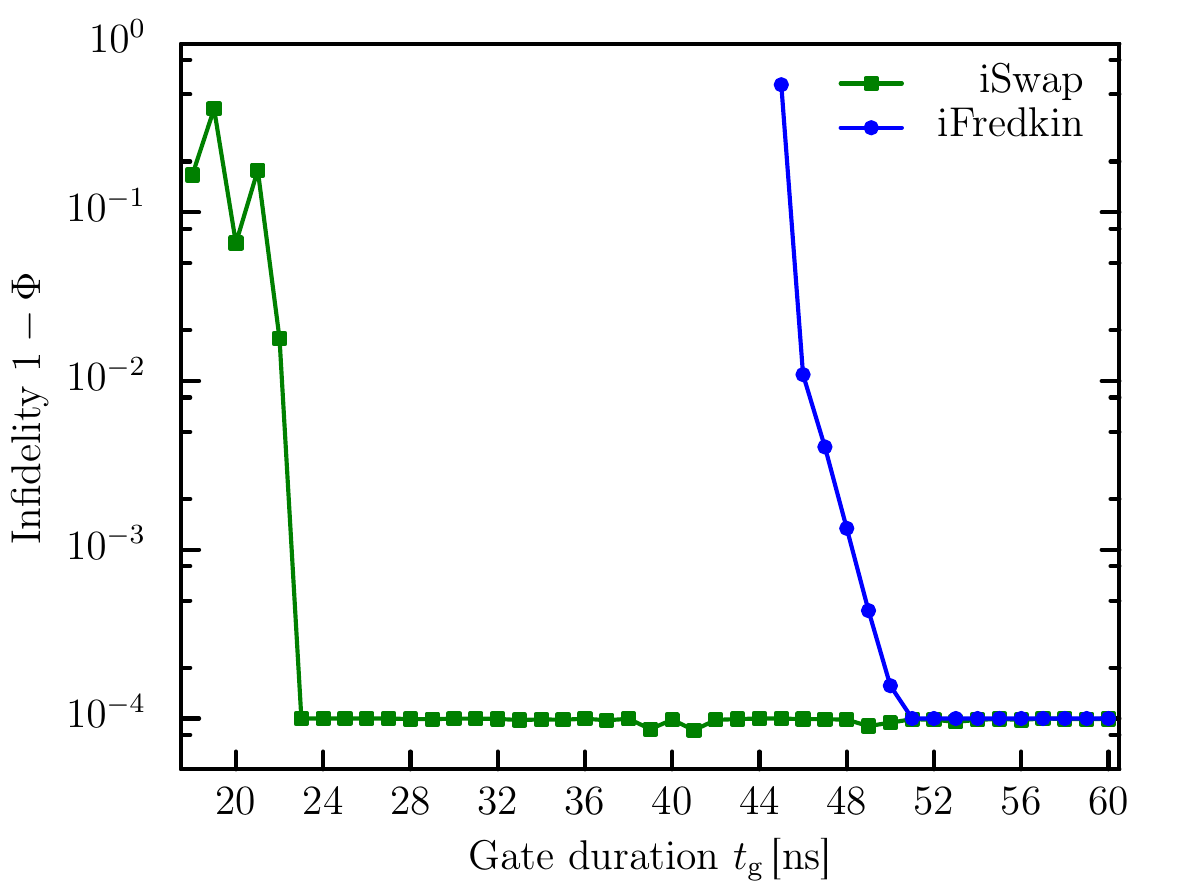}
  \caption{Speed limit for the 3 $Z$-controls for the fidelity $\Phi=0.9999$, compared to the iSWAP-gate with only two $Z$-controls, $S1$ ans $S2$, and $P$ is set to a parking frequency of 10\,GHz. The target fidelity in both cases is $\Phi=0.9999$.}
  \label{fig:sl}
\end{figure}

The following Hamiltonian describes the architecture, where the coupling between each qubit is mediated by a common bus
\begin{align}
  H =&
 \omega_B a^\dagger a + \sum_{i} \LR{\omega_i\LR{t} - \frac{\Delta_i}{2}} b_i^\dagger b_i + \frac{\Delta_i}{2} \LR{b_i^\dagger b_i}^2 \nonumber\\
  &+\sum_{i} g_i \LR{a^\dagger b_i + a b_i^\dagger}
 \,.
\end{align}
$a$ and $b_i$ are the bus and qubit annihilation operators, respectively. We pick realistic parameters for the architecture in order to proceed with the proof-of-principle. With a bus at $\omega_B/2\pi=6.5$\,GHz, the off-resonant frequencies are set to $\omega_P/2\pi=7.5$\,GHz, $\omega_{S1}/2\pi=8.0$\,GHz and $\omega_{S2}/2\pi=8.5$\,GHz, with anharmonicities $\Delta_P/2\pi=-200$\,MHz, $\Delta_{S1}/2\pi=-300$\,MHz and $\Delta_{S2}/2\pi=-400$\,MHz. The coupling strengths are $g_{BP}/2\pi=30$\,MHz, $g_{BS1}/2\pi=45$\,MHz and $g_{BS2}/2\pi=60$\,MHz, keeping the ratio $g_i/\Delta_i=-0.15$ fixed. In all runs the controls have a time resolution of $1\,$ns, typical for arbitrary waveform generators (AWGs), with fine steps of $0.1$\,ns for the simulated time evolution. Additionally, the pulse shapes are filtered by a Gaussian window with a bandwidth of 331\,MHz (standard deviation $\sigma=0.4$\,ns). For the optimization, we work in the rotating frame with angular frequency $\omega_R=\omega_B$. Therefore the implemented Hamiltonian reads
\begin{align}
H =&
  \sum_{i} \LR{\delta_i\LR{t} - \frac{\Delta_i}{2}} b_i^\dagger a_i + \frac{\Delta_i}{2} \LR{b_i^\dagger b_i}^2 \nonumber\\
  &+\sum_{i}  g^{(i)} \LR{a^\dagger b_i + a b_i^\dagger}
\,.
\end{align}
$\delta_i=\omega_i-\omega_B$ is the detuning of qubit $i$ from the bus. For each qubit, the first three energy levels are taken into account. Since we are only interested in the correct evolution of the computational subspace, the fidelity function only measures the overlap of the projected total time evolution with the target gate \cite{Rebentrost2009}
\begin{align}
\Phi = \frac{1}{4} \Abs{
  \Trace{U^\dagger_F P_Q U\LR{t_\mathrm{g}} P_Q}
}^2 \,,
\label{eq:phi}  
\end{align}
and global phases are omitted.

%

In fig.~\ref*{fig:zctrls} we show the optimized qubit-bus detuning parameter for the implementation of a iFREDKIN gate. The gate can be implemented in 56 ns for the chosen parameters with a realistic control sequence with $99.99$\% fidelity. The time evolution of the populations is shown in fig.~\ref{fig:pop}. As can be seen, the control qubit gets de-excited and re-excited during the process, hence allowing for the dynamics to be reenacted a two-excitation interference experiment, being consistent with the speed limit corresponding for a small multiple of the periods induced by the various $g$ couplings. Leakage into the second level of the qubits plays an important role in the gate implementation of the numerical pulse. Also, the pulse shapes are highly symmetric, as are the resulting time evolutions of the populations. The speed limit shown in fig.~\ref{fig:sl} proves that the iFREDKIN gate can easily be implemented below a gate duration of $t_\mathrm{g}=55$\,ns. This time scale is typical for analytic two-qubit pulse shapes, i.e., the simultaneous version of the Strauch sequence\cite{Strauch2003,Egger2014} and compares to the implementation of a traditional iSWAP on the same architecture as shown in fig.~\ref{fig:sl}, setting the $P$ qubit to a off-resonant parking frequency of $\omega_P/2\pi=10\,$GHz.

\section{\label{sec:conclusion}Conclusion} 
Quantum simulations could be one of the main applications for future quantum computers where they could outperform their classical counterparts.
We outlined the SPARQS circuit as an explicit superconducting implementation for a quantum simulator.
We showed how the most expensive gate in quantum simulations of Fermionic systems, the three-qubit iFREDKIN, can be efficiently implemented in such a device using GRAPE, a standard optimal control method.
For reasonable parameters of the SPARQS qubits, the iFREDKIN can be implemented in a time slightly longer than a typical iSWAP gate.
As coherence times for superconducting qubits keep increasing to date, it is realistic to expect that a large number of those gates could be reliably used for the purpose of performing quantum simulations beyond what can be done on classical computers.

\section*{acknowledgments}

We acknowledge support from the European Union under the ScaleQIT integrated project. 

\bibliography{literature} 

\begin{thebibliography}{30}%
\makeatletter
\providecommand \@ifxundefined [1]{%
 \@ifx{#1\undefined}
}%
\providecommand \@ifnum [1]{%
 \ifnum #1\expandafter \@firstoftwo
 \else \expandafter \@secondoftwo
 \fi
}%
\providecommand \@ifx [1]{%
 \ifx #1\expandafter \@firstoftwo
 \else \expandafter \@secondoftwo
 \fi
}%
\providecommand \natexlab [1]{#1}%
\providecommand \enquote  [1]{``#1''}%
\providecommand \bibnamefont  [1]{#1}%
\providecommand \bibfnamefont [1]{#1}%
\providecommand \citenamefont [1]{#1}%
\providecommand \href@noop [0]{\@secondoftwo}%
\providecommand \href [0]{\begingroup \@sanitize@url \@href}%
\providecommand \@href[1]{\@@startlink{#1}\@@href}%
\providecommand \@@href[1]{\endgroup#1\@@endlink}%
\providecommand \@sanitize@url [0]{\catcode `\\12\catcode `\$12\catcode
  `\&12\catcode `\#12\catcode `\^12\catcode `\_12\catcode `\%12\relax}%
\providecommand \@@startlink[1]{}%
\providecommand \@@endlink[0]{}%
\providecommand \url  [0]{\begingroup\@sanitize@url \@url }%
\providecommand \@url [1]{\endgroup\@href {#1}{\urlprefix }}%
\providecommand \urlprefix  [0]{URL }%
\providecommand \Eprint [0]{\href }%
\providecommand \doibase [0]{http://dx.doi.org/}%
\providecommand \selectlanguage [0]{\@gobble}%
\providecommand \bibinfo  [0]{\@secondoftwo}%
\providecommand \bibfield  [0]{\@secondoftwo}%
\providecommand \translation [1]{[#1]}%
\providecommand \BibitemOpen [0]{}%
\providecommand \bibitemStop [0]{}%
\providecommand \bibitemNoStop [0]{.\EOS\space}%
\providecommand \EOS [0]{\spacefactor3000\relax}%
\providecommand \BibitemShut  [1]{\csname bibitem#1\endcsname}%
\let\auto@bib@innerbib\@empty
\bibitem [{\citenamefont {Feynman}(1982)}]{Feynman1982}%
  \BibitemOpen
  \bibfield  {author} {\bibinfo {author} {\bibfnamefont {R.~P.}\ \bibnamefont
  {Feynman}},\ }\href {\doibase http://dx.doi.org/10.1007/BF02650179}
  {\bibfield  {journal} {\bibinfo  {journal} {International Journal of
  Theoretical Physics}\ }\textbf {\bibinfo {volume} {21}},\ \bibinfo {pages}
  {467} (\bibinfo {year} {1982})}\BibitemShut {NoStop}%
\bibitem [{\citenamefont {Shor}(1999)}]{Shor1999}%
  \BibitemOpen
  \bibfield  {author} {\bibinfo {author} {\bibfnamefont {P.~W.}\ \bibnamefont
  {Shor}},\ }\href {\doibase 10.1137/s0036144598347011} {\bibfield  {journal}
  {\bibinfo  {journal} {{SIAM} Rev.}\ }\textbf {\bibinfo {volume} {41}},\
  \bibinfo {pages} {303} (\bibinfo {year} {1999})}\BibitemShut {NoStop}%
\bibitem [{\citenamefont {DiVincenzo}(2000)}]{DiVincenzo2000}%
  \BibitemOpen
  \bibfield  {author} {\bibinfo {author} {\bibfnamefont {D.~P.}\ \bibnamefont
  {DiVincenzo}},\ }\href {\doibase
  http://dx.doi.org/10.1002/1521-3978(200009)48:9/11<771::AID-PROP771>3.0.CO;2-E}
  {\bibfield  {journal} {\bibinfo  {journal} {Fortschritte der Physik}\
  }\textbf {\bibinfo {volume} {48}},\ \bibinfo {pages} {771} (\bibinfo {year}
  {2000})}\BibitemShut {NoStop}%
\bibitem [{\citenamefont {Ladd}\ \emph {et~al.}(2010)\citenamefont {Ladd},
  \citenamefont {Jelezko}, \citenamefont {Laflamme}, \citenamefont {Nakamura},
  \citenamefont {Monroe},\ and\ \citenamefont {O'Brien}}]{Ladd2010}%
  \BibitemOpen
  \bibfield  {author} {\bibinfo {author} {\bibfnamefont {T.~D.}\ \bibnamefont
  {Ladd}}, \bibinfo {author} {\bibfnamefont {F.}~\bibnamefont {Jelezko}},
  \bibinfo {author} {\bibfnamefont {R.}~\bibnamefont {Laflamme}}, \bibinfo
  {author} {\bibfnamefont {Y.}~\bibnamefont {Nakamura}}, \bibinfo {author}
  {\bibfnamefont {C.}~\bibnamefont {Monroe}}, \ and\ \bibinfo {author}
  {\bibfnamefont {J.~L.}\ \bibnamefont {O'Brien}},\ }\href {\doibase
  http://dx.doi.org/10.1038/nature08812} {\bibfield  {journal} {\bibinfo
  {journal} {Nature}\ }\textbf {\bibinfo {volume} {464}},\ \bibinfo {pages}
  {45} (\bibinfo {year} {2010})}\BibitemShut {NoStop}%
\bibitem [{\citenamefont {Fujii}(2015)}]{Fujii2015}%
  \BibitemOpen
  \bibfield  {author} {\bibinfo {author} {\bibfnamefont {K.}~\bibnamefont
  {Fujii}},\ }\href {\doibase http://dx.doi.org/10.1007/978-981-287-996-7}
  {\emph {\bibinfo {title} {Quantum Computation with Topological Codes}}}\
  (\bibinfo  {publisher} {Springer Singapore},\ \bibinfo {year}
  {2015})\BibitemShut {NoStop}%
\bibitem [{\citenamefont {Bloch}\ \emph {et~al.}(2012)\citenamefont {Bloch},
  \citenamefont {Dalibard},\ and\ \citenamefont
  {Nascimb{\`{e}}ne}}]{Bloch2012}%
  \BibitemOpen
  \bibfield  {author} {\bibinfo {author} {\bibfnamefont {I.}~\bibnamefont
  {Bloch}}, \bibinfo {author} {\bibfnamefont {J.}~\bibnamefont {Dalibard}}, \
  and\ \bibinfo {author} {\bibfnamefont {S.}~\bibnamefont {Nascimb{\`{e}}ne}},\
  }\href {\doibase http://dx.doi.org/10.1038/nphys2259} {\bibfield  {journal}
  {\bibinfo  {journal} {Nature Physics}\ }\textbf {\bibinfo {volume} {8}},\
  \bibinfo {pages} {267} (\bibinfo {year} {2012})}\BibitemShut {NoStop}%
\bibitem [{\citenamefont {Blatt}\ and\ \citenamefont {Roos}(2012)}]{Blatt2012}%
  \BibitemOpen
  \bibfield  {author} {\bibinfo {author} {\bibfnamefont {R.}~\bibnamefont
  {Blatt}}\ and\ \bibinfo {author} {\bibfnamefont {C.~F.}\ \bibnamefont
  {Roos}},\ }\href {\doibase http://dx.doi.org/10.1038/nphys2252} {\bibfield
  {journal} {\bibinfo  {journal} {Nature Physics}\ }\textbf {\bibinfo {volume}
  {8}},\ \bibinfo {pages} {277} (\bibinfo {year} {2012})}\BibitemShut {NoStop}%
\bibitem [{\citenamefont {Aspuru-Guzik}\ and\ \citenamefont
  {Walther}(2012)}]{AspuruGuzik2012}%
  \BibitemOpen
  \bibfield  {author} {\bibinfo {author} {\bibfnamefont {A.}~\bibnamefont
  {Aspuru-Guzik}}\ and\ \bibinfo {author} {\bibfnamefont {P.}~\bibnamefont
  {Walther}},\ }\href {\doibase http://dx.doi.org/10.1038/nphys2253} {\bibfield
   {journal} {\bibinfo  {journal} {Nature Physics}\ }\textbf {\bibinfo {volume}
  {8}},\ \bibinfo {pages} {285} (\bibinfo {year} {2012})}\BibitemShut {NoStop}%
\bibitem [{\citenamefont {Houck}\ \emph {et~al.}(2012)\citenamefont {Houck},
  \citenamefont {Türeci},\ and\ \citenamefont {Koch}}]{Houck2012}%
  \BibitemOpen
  \bibfield  {author} {\bibinfo {author} {\bibfnamefont {A.~A.}\ \bibnamefont
  {Houck}}, \bibinfo {author} {\bibfnamefont {H.~E.}\ \bibnamefont {Türeci}},
  \ and\ \bibinfo {author} {\bibfnamefont {J.}~\bibnamefont {Koch}},\ }\href
  {\doibase http://dx.doi.org/10.1038/nphys2251} {\bibfield  {journal}
  {\bibinfo  {journal} {Nature Physics}\ }\textbf {\bibinfo {volume} {8}},\
  \bibinfo {pages} {292} (\bibinfo {year} {2012})}\BibitemShut {NoStop}%
\bibitem [{\citenamefont {Georgescu}\ \emph {et~al.}(2014)\citenamefont
  {Georgescu}, \citenamefont {Ashhab},\ and\ \citenamefont
  {Nori}}]{Georgescu2014}%
  \BibitemOpen
  \bibfield  {author} {\bibinfo {author} {\bibfnamefont {I.~M.}\ \bibnamefont
  {Georgescu}}, \bibinfo {author} {\bibfnamefont {S.}~\bibnamefont {Ashhab}}, \
  and\ \bibinfo {author} {\bibfnamefont {F.}~\bibnamefont {Nori}},\ }\href
  {\doibase https://doi.org/10.1103/RevModPhys.86.153} {\bibfield  {journal}
  {\bibinfo  {journal} {Reviews of Modern Physics}\ }\textbf {\bibinfo {volume}
  {86}},\ \bibinfo {pages} {153} (\bibinfo {year} {2014})}\BibitemShut
  {NoStop}%
\bibitem [{\citenamefont {Mariantoni}\ \emph {et~al.}(2011)\citenamefont
  {Mariantoni}, \citenamefont {Wang}, \citenamefont {Yamamoto}, \citenamefont
  {Neeley}, \citenamefont {Bialczak}, \citenamefont {Chen}, \citenamefont
  {Lenander}, \citenamefont {Lucero}, \citenamefont
  {O{\textquoteright}Connell}, \citenamefont {Sank}, \citenamefont {Weides},
  \citenamefont {Wenner}, \citenamefont {Yin}, \citenamefont {Zhao},
  \citenamefont {Korotkov}, \citenamefont {Cleland},\ and\ \citenamefont
  {Martinis}}]{Mariantoni2011}%
  \BibitemOpen
  \bibfield  {author} {\bibinfo {author} {\bibfnamefont {M.}~\bibnamefont
  {Mariantoni}}, \bibinfo {author} {\bibfnamefont {H.}~\bibnamefont {Wang}},
  \bibinfo {author} {\bibfnamefont {T.}~\bibnamefont {Yamamoto}}, \bibinfo
  {author} {\bibfnamefont {M.}~\bibnamefont {Neeley}}, \bibinfo {author}
  {\bibfnamefont {R.~C.}\ \bibnamefont {Bialczak}}, \bibinfo {author}
  {\bibfnamefont {Y.}~\bibnamefont {Chen}}, \bibinfo {author} {\bibfnamefont
  {M.}~\bibnamefont {Lenander}}, \bibinfo {author} {\bibfnamefont
  {E.}~\bibnamefont {Lucero}}, \bibinfo {author} {\bibfnamefont {A.~D.}\
  \bibnamefont {O{\textquoteright}Connell}}, \bibinfo {author} {\bibfnamefont
  {D.}~\bibnamefont {Sank}}, \bibinfo {author} {\bibfnamefont {M.}~\bibnamefont
  {Weides}}, \bibinfo {author} {\bibfnamefont {J.}~\bibnamefont {Wenner}},
  \bibinfo {author} {\bibfnamefont {Y.}~\bibnamefont {Yin}}, \bibinfo {author}
  {\bibfnamefont {J.}~\bibnamefont {Zhao}}, \bibinfo {author} {\bibfnamefont
  {A.~N.}\ \bibnamefont {Korotkov}}, \bibinfo {author} {\bibfnamefont {A.~N.}\
  \bibnamefont {Cleland}}, \ and\ \bibinfo {author} {\bibfnamefont {J.~M.}\
  \bibnamefont {Martinis}},\ }\href {\doibase 10.1126/science.1208517}
  {\bibfield  {journal} {\bibinfo  {journal} {Science}\ }\textbf {\bibinfo
  {volume} {334}},\ \bibinfo {pages} {61} (\bibinfo {year} {2011})}\BibitemShut
  {NoStop}%
\bibitem [{\citenamefont {Galiautdinov}\ \emph {et~al.}(2012)\citenamefont
  {Galiautdinov}, \citenamefont {Korotkov},\ and\ \citenamefont
  {Martinis}}]{Galiautdinov2012}%
  \BibitemOpen
  \bibfield  {author} {\bibinfo {author} {\bibfnamefont {A.}~\bibnamefont
  {Galiautdinov}}, \bibinfo {author} {\bibfnamefont {A.~N.}\ \bibnamefont
  {Korotkov}}, \ and\ \bibinfo {author} {\bibfnamefont {J.~M.}\ \bibnamefont
  {Martinis}},\ }\href@noop {} {\bibfield  {journal} {\bibinfo  {journal}
  {Phys. Rev. A}\ }\textbf {\bibinfo {volume} {85}} (\bibinfo {year}
  {2012})}\BibitemShut {NoStop}%
\bibitem [{\citenamefont {Ghosh}\ \emph {et~al.}(2013)\citenamefont {Ghosh},
  \citenamefont {Galiautdinov}, \citenamefont {Zhou}, \citenamefont {Korotkov},
  \citenamefont {Martinis},\ and\ \citenamefont {Geller}}]{Ghosh2013}%
  \BibitemOpen
  \bibfield  {author} {\bibinfo {author} {\bibfnamefont {J.}~\bibnamefont
  {Ghosh}}, \bibinfo {author} {\bibfnamefont {A.}~\bibnamefont {Galiautdinov}},
  \bibinfo {author} {\bibfnamefont {Z.}~\bibnamefont {Zhou}}, \bibinfo {author}
  {\bibfnamefont {A.~N.}\ \bibnamefont {Korotkov}}, \bibinfo {author}
  {\bibfnamefont {J.~M.}\ \bibnamefont {Martinis}}, \ and\ \bibinfo {author}
  {\bibfnamefont {M.~R.}\ \bibnamefont {Geller}},\ }\href {\doibase
  10.1103/PhysRevA.87.022309} {\bibfield  {journal} {\bibinfo  {journal} {Phys.
  Rev. A}\ }\textbf {\bibinfo {volume} {87}},\ \bibinfo {pages} {022309}
  (\bibinfo {year} {2013})}\BibitemShut {NoStop}%
\bibitem [{\citenamefont {Dallaire-Demers}\ and\ \citenamefont
  {Wilhelm}(2016{\natexlab{a}})}]{DDW16}%
  \BibitemOpen
  \bibfield  {author} {\bibinfo {author} {\bibfnamefont {P.-L.}\ \bibnamefont
  {Dallaire-Demers}}\ and\ \bibinfo {author} {\bibfnamefont {F.~K.}\
  \bibnamefont {Wilhelm}},\ }\href {\doibase
  http://dx.doi.org/10.1103/PhysRevA.93.032303} {\bibfield  {journal} {\bibinfo
   {journal} {Phys. Rev. A}\ }\textbf {\bibinfo {volume} {93}},\ \bibinfo
  {pages} {032303} (\bibinfo {year} {2016}{\natexlab{a}})}\BibitemShut
  {NoStop}%
\bibitem [{\citenamefont {Dallaire-Demers}\ and\ \citenamefont
  {Wilhelm}(2016{\natexlab{b}})}]{DDW16b}%
  \BibitemOpen
  \bibfield  {author} {\bibinfo {author} {\bibfnamefont {P.-L.}\ \bibnamefont
  {Dallaire-Demers}}\ and\ \bibinfo {author} {\bibfnamefont {F.~K.}\
  \bibnamefont {Wilhelm}},\ }\href {\doibase
  https://doi.org/10.1103/PhysRevA.94.062304} {\bibfield  {journal} {\bibinfo
  {journal} {Phys. Rev. A}\ }\textbf {\bibinfo {volume} {94}},\ \bibinfo
  {pages} {062304} (\bibinfo {year} {2016}{\natexlab{b}})}\BibitemShut
  {NoStop}%
\bibitem [{\citenamefont {Egger}\ and\ \citenamefont
  {Wilhelm}(2014)}]{Egger2014}%
  \BibitemOpen
  \bibfield  {author} {\bibinfo {author} {\bibfnamefont {D.~J.}\ \bibnamefont
  {Egger}}\ and\ \bibinfo {author} {\bibfnamefont {F.~K.}\ \bibnamefont
  {Wilhelm}},\ }\href {\doibase 10.1088/0953-2048/27/1/014001} {\bibfield
  {journal} {\bibinfo  {journal} {Supercond. Sci. Technol.}\ }\textbf {\bibinfo
  {volume} {27}},\ \bibinfo {pages} {014001} (\bibinfo {year}
  {2014})}\BibitemShut {NoStop}%
\bibitem [{\citenamefont {Calderon-Vargas}\ and\ \citenamefont
  {Kestner}(2015)}]{Calderon-Vargas2015}%
  \BibitemOpen
  \bibfield  {author} {\bibinfo {author} {\bibfnamefont {F.~A.}\ \bibnamefont
  {Calderon-Vargas}}\ and\ \bibinfo {author} {\bibfnamefont {J.~P.}\
  \bibnamefont {Kestner}},\ }\href {\doibase 10.1103/PhysRevB.91.035301}
  {\bibfield  {journal} {\bibinfo  {journal} {Phys. Rev. B}\ }\textbf {\bibinfo
  {volume} {91}},\ \bibinfo {pages} {035301} (\bibinfo {year}
  {2015})}\BibitemShut {NoStop}%
\bibitem [{\citenamefont {Khaneja}\ \emph {et~al.}(2005)\citenamefont
  {Khaneja}, \citenamefont {Reiss}, \citenamefont {Kehlet}, \citenamefont
  {Schulte-{H}erbr\"{u}ggen},\ and\ \citenamefont {Glaser}}]{Khaneja2005}%
  \BibitemOpen
  \bibfield  {author} {\bibinfo {author} {\bibfnamefont {N.}~\bibnamefont
  {Khaneja}}, \bibinfo {author} {\bibfnamefont {T.}~\bibnamefont {Reiss}},
  \bibinfo {author} {\bibfnamefont {C.}~\bibnamefont {Kehlet}}, \bibinfo
  {author} {\bibfnamefont {T.}~\bibnamefont {Schulte-{H}erbr\"{u}ggen}}, \ and\
  \bibinfo {author} {\bibfnamefont {S.~J.}\ \bibnamefont {Glaser}},\ }\href
  {\doibase 10.1016/j.jmr.2004.11.004} {\bibfield  {journal} {\bibinfo
  {journal} {J. Magn. Reson.}\ }\textbf {\bibinfo {volume} {172}},\ \bibinfo
  {pages} {296} (\bibinfo {year} {2005})}\BibitemShut {NoStop}%
\bibitem [{\citenamefont {Glaser}\ \emph {et~al.}(2015)\citenamefont {Glaser},
  \citenamefont {Boscain}, \citenamefont {Calarco}, \citenamefont {Koch},
  \citenamefont {K\"ockenberger}, \citenamefont {Kosloff}, \citenamefont
  {Kuprov}, \citenamefont {Luy}, \citenamefont {Schirmer}, \citenamefont
  {Schulte-{H}erbr\"uggen}, \citenamefont {Sugny},\ and\ \citenamefont
  {Wilhelm}}]{Glaser2015}%
  \BibitemOpen
  \bibfield  {author} {\bibinfo {author} {\bibfnamefont {S.~J.}\ \bibnamefont
  {Glaser}}, \bibinfo {author} {\bibfnamefont {U.}~\bibnamefont {Boscain}},
  \bibinfo {author} {\bibfnamefont {T.}~\bibnamefont {Calarco}}, \bibinfo
  {author} {\bibfnamefont {C.~P.}\ \bibnamefont {Koch}}, \bibinfo {author}
  {\bibfnamefont {W.}~\bibnamefont {K\"ockenberger}}, \bibinfo {author}
  {\bibfnamefont {R.}~\bibnamefont {Kosloff}}, \bibinfo {author} {\bibfnamefont
  {I.}~\bibnamefont {Kuprov}}, \bibinfo {author} {\bibfnamefont
  {B.}~\bibnamefont {Luy}}, \bibinfo {author} {\bibfnamefont {S.}~\bibnamefont
  {Schirmer}}, \bibinfo {author} {\bibfnamefont {T.}~\bibnamefont
  {Schulte-{H}erbr\"uggen}}, \bibinfo {author} {\bibfnamefont {D.}~\bibnamefont
  {Sugny}}, \ and\ \bibinfo {author} {\bibfnamefont {F.~K.}\ \bibnamefont
  {Wilhelm}},\ }\href {\doibase 10.1140/epjd/e2015-60464-1} {\bibfield
  {journal} {\bibinfo  {journal} {Eur. Phys. J. D}\ }\textbf {\bibinfo {volume}
  {69}},\ \bibinfo {pages} {279} (\bibinfo {year} {2015})}\BibitemShut
  {NoStop}%
\bibitem [{\citenamefont {Riera}\ \emph {et~al.}(2012)\citenamefont {Riera},
  \citenamefont {Gogolin},\ and\ \citenamefont {Eisert}}]{Riera2012}%
  \BibitemOpen
  \bibfield  {author} {\bibinfo {author} {\bibfnamefont {A.}~\bibnamefont
  {Riera}}, \bibinfo {author} {\bibfnamefont {C.}~\bibnamefont {Gogolin}}, \
  and\ \bibinfo {author} {\bibfnamefont {J.}~\bibnamefont {Eisert}},\ }\href
  {\doibase http://dx.doi.org/10.1103/PhysRevLett.108.080402} {\bibfield
  {journal} {\bibinfo  {journal} {Phys. Rev. Lett.}\ }\textbf {\bibinfo
  {volume} {108}},\ \bibinfo {pages} {080402} (\bibinfo {year}
  {2012})}\BibitemShut {NoStop}%
\bibitem [{\citenamefont {Nielsen}\ and\ \citenamefont
  {Chuang}(2010)}]{Nielsen2010book}%
  \BibitemOpen
  \bibfield  {author} {\bibinfo {author} {\bibfnamefont {M.}~\bibnamefont
  {Nielsen}}\ and\ \bibinfo {author} {\bibfnamefont {I.}~\bibnamefont
  {Chuang}},\ }\href@noop {} {\emph {\bibinfo {title} {Quantum Computation and
  Quantum Information}}}\ (\bibinfo  {publisher} {Cambridge University Press},\
  \bibinfo {year} {2010})\BibitemShut {NoStop}%
\bibitem [{\citenamefont {Blais}\ \emph {et~al.}(2004)\citenamefont {Blais},
  \citenamefont {Huang}, \citenamefont {Wallraff}, \citenamefont {Girvin},\
  and\ \citenamefont {Schoelkopf}}]{Blais2004}%
  \BibitemOpen
  \bibfield  {author} {\bibinfo {author} {\bibfnamefont {A.}~\bibnamefont
  {Blais}}, \bibinfo {author} {\bibfnamefont {R.-S.}\ \bibnamefont {Huang}},
  \bibinfo {author} {\bibfnamefont {A.}~\bibnamefont {Wallraff}}, \bibinfo
  {author} {\bibfnamefont {S.~M.}\ \bibnamefont {Girvin}}, \ and\ \bibinfo
  {author} {\bibfnamefont {R.~J.}\ \bibnamefont {Schoelkopf}},\ }\href
  {\doibase 10.1103/PhysRevA.69.062320} {\bibfield  {journal} {\bibinfo
  {journal} {Phys. Rev. A}\ }\textbf {\bibinfo {volume} {69}},\ \bibinfo
  {pages} {062320} (\bibinfo {year} {2004})}\BibitemShut {NoStop}%
\bibitem [{\citenamefont {Blais}\ \emph {et~al.}(2007)\citenamefont {Blais},
  \citenamefont {Gambetta}, \citenamefont {Wallraff}, \citenamefont {Schuster},
  \citenamefont {Girvin}, \citenamefont {Devoret},\ and\ \citenamefont
  {Schoelkopf}}]{Blais2007}%
  \BibitemOpen
  \bibfield  {author} {\bibinfo {author} {\bibfnamefont {A.}~\bibnamefont
  {Blais}}, \bibinfo {author} {\bibfnamefont {J.}~\bibnamefont {Gambetta}},
  \bibinfo {author} {\bibfnamefont {A.}~\bibnamefont {Wallraff}}, \bibinfo
  {author} {\bibfnamefont {D.~I.}\ \bibnamefont {Schuster}}, \bibinfo {author}
  {\bibfnamefont {S.~M.}\ \bibnamefont {Girvin}}, \bibinfo {author}
  {\bibfnamefont {M.~H.}\ \bibnamefont {Devoret}}, \ and\ \bibinfo {author}
  {\bibfnamefont {R.~J.}\ \bibnamefont {Schoelkopf}},\ }\href {\doibase
  10.1103/PhysRevA.75.032329} {\bibfield  {journal} {\bibinfo  {journal} {Phys.
  Rev. A}\ }\textbf {\bibinfo {volume} {75}},\ \bibinfo {pages} {032329}
  (\bibinfo {year} {2007})}\BibitemShut {NoStop}%
\bibitem [{\citenamefont {Geerlings}(2013)}]{Geerlings13}%
  \BibitemOpen
  \bibfield  {author} {\bibinfo {author} {\bibfnamefont {K.~L.}\ \bibnamefont
  {Geerlings}},\ }\href@noop {} {Ph.D. thesis},\ \bibinfo  {school} {Yale
  University} (\bibinfo {year} {2013})\BibitemShut {NoStop}%
\bibitem [{\citenamefont {Jordan}\ and\ \citenamefont
  {Wigner}(1928)}]{Jordan1928}%
  \BibitemOpen
  \bibfield  {author} {\bibinfo {author} {\bibfnamefont {P.}~\bibnamefont
  {Jordan}}\ and\ \bibinfo {author} {\bibfnamefont {E.}~\bibnamefont
  {Wigner}},\ }\href {\doibase http://dx.doi.org/10.1007/BF01331938} {\bibfield
   {journal} {\bibinfo  {journal} {Z. Phys.}\ }\textbf {\bibinfo {volume}
  {47}},\ \bibinfo {pages} {631} (\bibinfo {year} {1928})}\BibitemShut
  {NoStop}%
\bibitem [{\citenamefont {D\"ur}\ \emph {et~al.}(2000)\citenamefont {D\"ur},
  \citenamefont {Vidal},\ and\ \citenamefont {Cirac}}]{Dur2000}%
  \BibitemOpen
  \bibfield  {author} {\bibinfo {author} {\bibfnamefont {W.}~\bibnamefont
  {D\"ur}}, \bibinfo {author} {\bibfnamefont {G.}~\bibnamefont {Vidal}}, \ and\
  \bibinfo {author} {\bibfnamefont {J.~I.}\ \bibnamefont {Cirac}},\ }\href
  {\doibase http://dx.doi.org/10.1103/PhysRevA.62.062314} {\bibfield  {journal}
  {\bibinfo  {journal} {Phys. Rev. A}\ }\textbf {\bibinfo {volume} {62}},\
  \bibinfo {pages} {062314} (\bibinfo {year} {2000})}\BibitemShut {NoStop}%
\bibitem [{\citenamefont {Zhang}\ \emph {et~al.}(2003)\citenamefont {Zhang},
  \citenamefont {Vala}, \citenamefont {Sastry},\ and\ \citenamefont
  {Whaley}}]{Zhang2003}%
  \BibitemOpen
  \bibfield  {author} {\bibinfo {author} {\bibfnamefont {J.}~\bibnamefont
  {Zhang}}, \bibinfo {author} {\bibfnamefont {J.}~\bibnamefont {Vala}},
  \bibinfo {author} {\bibfnamefont {S.}~\bibnamefont {Sastry}}, \ and\ \bibinfo
  {author} {\bibfnamefont {K.~B.}\ \bibnamefont {Whaley}},\ }\href@noop {}
  {\bibfield  {journal} {\bibinfo  {journal} {Physical Review A}\ }\textbf
  {\bibinfo {volume} {67}} (\bibinfo {year} {2003})}\BibitemShut {NoStop}%
\bibitem [{\citenamefont {Zahedinejad}\ \emph {et~al.}(2015)\citenamefont
  {Zahedinejad}, \citenamefont {Ghosh},\ and\ \citenamefont
  {Sanders}}]{Zahedinejad2015}%
  \BibitemOpen
  \bibfield  {author} {\bibinfo {author} {\bibfnamefont {E.}~\bibnamefont
  {Zahedinejad}}, \bibinfo {author} {\bibfnamefont {J.}~\bibnamefont {Ghosh}},
  \ and\ \bibinfo {author} {\bibfnamefont {B.~C.}\ \bibnamefont {Sanders}},\
  }\href {\doibase 10.1103/PhysRevLett.114.200502} {\bibfield  {journal}
  {\bibinfo  {journal} {Phys. Rev. Lett.}\ }\textbf {\bibinfo {volume} {114}},\
  \bibinfo {pages} {200502} (\bibinfo {year} {2015})}\BibitemShut {NoStop}%
\bibitem [{\citenamefont {Rebentrost}\ and\ \citenamefont
  {Wilhelm}(2009)}]{Rebentrost2009}%
  \BibitemOpen
  \bibfield  {author} {\bibinfo {author} {\bibfnamefont {P.}~\bibnamefont
  {Rebentrost}}\ and\ \bibinfo {author} {\bibfnamefont {F.~K.}\ \bibnamefont
  {Wilhelm}},\ }\href {\doibase 10.1103/PhysRevB.79.060507} {\bibfield
  {journal} {\bibinfo  {journal} {Phys. Rev. B}\ }\textbf {\bibinfo {volume}
  {79}},\ \bibinfo {pages} {060507} (\bibinfo {year} {2009})}\BibitemShut
  {NoStop}%
\bibitem [{\citenamefont {Strauch}\ \emph {et~al.}(2003)\citenamefont
  {Strauch}, \citenamefont {Johnson}, \citenamefont {Dragt}, \citenamefont
  {Lobb}, \citenamefont {Anderson},\ and\ \citenamefont
  {Wellstood}}]{Strauch2003}%
  \BibitemOpen
  \bibfield  {author} {\bibinfo {author} {\bibfnamefont {F.~W.}\ \bibnamefont
  {Strauch}}, \bibinfo {author} {\bibfnamefont {P.~R.}\ \bibnamefont
  {Johnson}}, \bibinfo {author} {\bibfnamefont {A.~J.}\ \bibnamefont {Dragt}},
  \bibinfo {author} {\bibfnamefont {C.~J.}\ \bibnamefont {Lobb}}, \bibinfo
  {author} {\bibfnamefont {J.~R.}\ \bibnamefont {Anderson}}, \ and\ \bibinfo
  {author} {\bibfnamefont {F.~C.}\ \bibnamefont {Wellstood}},\ }\href {\doibase
  10.1103/PhysRevLett.91.167005} {\bibfield  {journal} {\bibinfo  {journal}
  {Phys. Rev. Lett.}\ }\textbf {\bibinfo {volume} {91}},\ \bibinfo {pages}
  {167005} (\bibinfo {year} {2003})}\BibitemShut {NoStop}%
\end{thebibliography}%

\end{document}